\documentclass[aps,prb,twocolumn,superscriptaddress,longbibliography]{revtex4-2}
\usepackage{epsfig,amsopn}
\usepackage{graphicx}
\usepackage{color}
\usepackage{amsmath,amssymb}
\usepackage{enumerate}
\newcommand\bea{\begin{eqnarray}}
\newcommand\eea{\end{eqnarray}}
\newcommand\beq{\begin{equation}}
\newcommand\eeq{\end{equation}}

\def\nn{\nonumber}
\def\f{\frac}
\def\al{\alpha}

\def\ep{\epsilon}

\def\si{\sigma}

\def\dg{\dagger}

\def\ua{\uparrow}
\def\da{\downarrow}

\def\th{\theta}

\begin{document}
\title{Persistent currents in mesoscopic spin-orbit coupled rings due to an applied Zeeman field}
\author{Bijay Kumar Sahoo}
\affiliation{School of Physics, University of Hyderabad, Prof. C. R. Rao Road, Gachibowli, Hyderabad-500046, India}
\author{Subroto Mukerjee}
\affiliation{Department of Physics, Indian Institute of Science, Bengaluru-560012, India}
\author{ Abhiram Soori}
\email{abhirams@uohyd.ac.in}
\affiliation{School of Physics, University of Hyderabad, Prof. C. R. Rao Road, Gachibowli, Hyderabad-500046, India}
\begin{abstract}
Persistent currents (PCs) in mesoscopic rings have been a subject of intense investigation since their proposal by B\"uttiker, Landauer, and Imry in 1983. In this paper, we explore the behavior of PC in spin-orbit coupled rings under the influence of a Zeeman field (without a need for a flux threading the ring), contrasting it with traditional PC observed in rings threaded by magnetic flux. Our study reveals that the emergence of PC in our setup crucially depends on nonzero values of spin-orbit coupling and the Zeeman field. Through theoretical analysis and  numerical calculations, we uncover several intriguing phenomena. Specifically, in ballistic rings, we observe an inverse proportionality between PC and system size, with PC being zero at half filling for even numbers of sites. Additionally, the introduction of on-site disorder leads to the suppression of PC, with exponential decay observed for large disorder strengths and quadratic decay for smaller disorder strengths. Notably, disorder can enhance PC in individual samples, albeit with a configuration-averaged PC of zero. Furthermore, we find that the standard deviation of PC increases with disorder strength, reaching a maximum before decreasing to zero at high disorder strengths. We study the case of PC when the Zeeman field and the spin-orbit field are noncollinear. We also study persistent spin current which shows behavior similar to that of PC except that at half filling, it is not zero.  Our findings shed light on the intricate interplay between spin-orbit coupling, Zeeman fields, and disorder in mesoscopic quantum systems, offering new avenues for theoretical exploration and experimental verification.
\end{abstract}

\maketitle
\section{Introduction}

In 1983, Buttiker, Landauer, and Imry introduced the concept of undecaying current in non-superconducting metallic rings threaded by a magnetic flux~\cite{BUTTIKER1983365}. This persistent current (PC) phenomenon manifests when the temperature is sufficiently low ($<1K$) and the ring size is small~\cite{viewpoint1.7} ($<1\mu m$). In this realm, the energy levels of closed systems become discrete~\cite{Cheung}, and interference effects become pivotal when the system size is comparable to or less than the phase coherence length~\cite{Cheung}, $L_\phi$. These PCs exhibit periodic behavior with respect to magnetic flux, with a period equal to the flux quantum~\cite{Cheung} $\phi_0 = h/e$. Remarkably, even static disorder fails to obliterate them~\cite{BUTTIKER1983365}. For weak disorder, the PC ($I$) varies quadratically with the disorder strength ($w$), while for strong disorder, it decays exponentially with the ring size~\cite{Cheung}. Also, with number of channels, PC increases and in realistic multichannel rings, the PC was  predicted to be observable~\cite{riedel1989}. PC in corbino geometry has also been investigated theoretically~\cite{yerin2021}. An explanation for the experimentally observed PC in normal metals concerning magnetic impurities and attractive interactions was provided by H. Bary-Soroker et al. in 2008~\cite{Bary-Soroker2008}.

The theoretical exploration of PCs has been extensive for more than two decades. However, its experimental detection poses significant challenges due to the minute signal produced and its high sensitivity to the environment. The first experimental observation of PC dates back to nearly three decades ago, employing SQUID techniques~\cite{Levy1990, Chandrasekhar1991, Maily1993}. However, discrepancies with theoretical predictions were noted. Subsequent measurements utilizing micromechanical detectors overcame these limitations, providing enhanced sensitivity and reduced back action~\cite{P}. Moreover, these experimental results aligned well with theoretical expectations, enabling more comprehensive investigations into PC dynamics across single rings and ring arrays concerning size, temperature, and magnetic field orientation.

While  Hall effect, a transverse voltage arising in response to a longitudinal current in a two-dimensional metal under a normal magnetic field, is well known,  planar Hall effect occurs in samples with spin-orbit coupling under the influence  of an in-plane magnetic field~\cite{kittel,goldberg54,tang03,Roy10,annadi13,taskin17}. The Hall effect is due to Lorentz force on electrons whereas the planar Hall is due to a combination of Zeeman field and spin-orbit coupling~\cite{suri21,soori2021}. In Datta-Das spin transistors, transverse current is generated (in systems with periodic boundary conditions in transverse direction) due to spin-polarized electron injection into a spin-orbit coupled central region, instead of an in-plane Zeeman field~\cite{Sahoo2023}. An implication of this work is the persistence of nonzero transverse conductivity even when the junction is cut off, attributable to the combined influence of injected spin-polarized electrons and spin-orbit coupling in the central region. Zero longitudinal current suggests that the system is in equilibrium. A nonzero transverse current in an equilibrium setup  prompts the query: ``does a PC flow in an isolated spin-orbit coupled ring under the influence of a Zeeman field?". 
 
 In Sec. II, we perform calculation of PC with no disorder in the system and check its behaviour with electron filling,  ($\al$, $b$), and the size of the ring. In Sec. III, we discuss the effects of the disorder on  PC and how it varies with the strength of the disorder for experimentally relevant values of  ($\al$, $b$). These values are computed from the estimated data of InAs nanowires~\cite{InAs2010}.  We find that at half-filling  PC is exactly zero in the ballistic case, but with the disorder, we can have a nonzero PC for a fixed disorder configuration, though the configuration averaged PC is zero.

\section{Ballistic Ring}
 
 The system can be described as a lattice model by the following Hamiltonian 
\bea 
\label{eq:Ham}
H&=& \sum_{n=1}^{N}\Big [-t(c_{n+1}^\dg c_n+h.c)+\f{\al}{2}(ic_{n+1}^\dg \si_zc_n+{\rm h.c.})\nn \\ &&~~~~~~~+bc_{n}^\dg \si_{\th,\phi}c_n \Big] \eea
where $c_n =\begin{bmatrix}
                      c_{n,\ua} & c_{n,\da}\\                      
       \end{bmatrix}$$^T$, $c_{n,\si}$ annihilates an electron of spin $\si$ at site $n$. The first term in the Hamiltonian represents a tight binding term with $t$ being the hopping strength, the second term corresponds to spin-orbit coupling with $\al$ as  coupling strength, and the last term contains the Zeeman field with $b$ being the Zeeman energy. Here, $\si_{\th,\phi}=\cos{\th}\si_z+\sin{\th}(\cos{\phi}\si_x+\sin{\phi}\si_y)$, with $\si_{j}$ and $j=x,y,z$ being the Pauli spin matrices. $N$ is the number of sites in the ring. The coordinates $(\th,\phi)$ determine the direction of the Zeeman field applied. Also, $N+1\equiv 1$ and the system is periodic. So, momentum $\hbar k$ is a good quantum number and the wave number $k$ takes values ($2\pi j/N$) for $j=0,1,2,..,N-1$. The ring is assumed to be in the $xy$-plane, while the spin orbit field is directed along $\hat z$-axis. 
\begin{figure}[htb]
 \includegraphics[width=4cm]{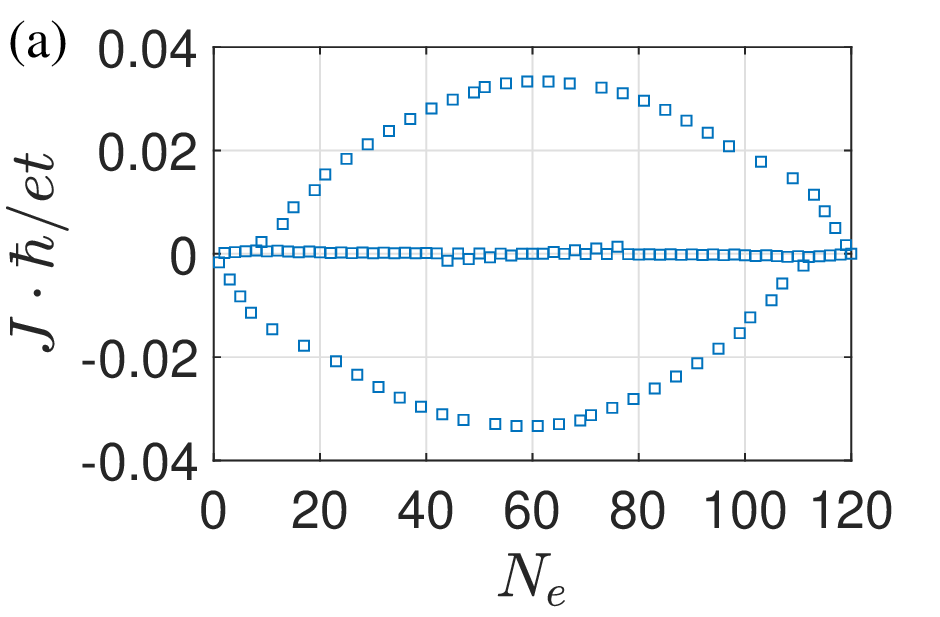}
 \includegraphics[width=4cm]{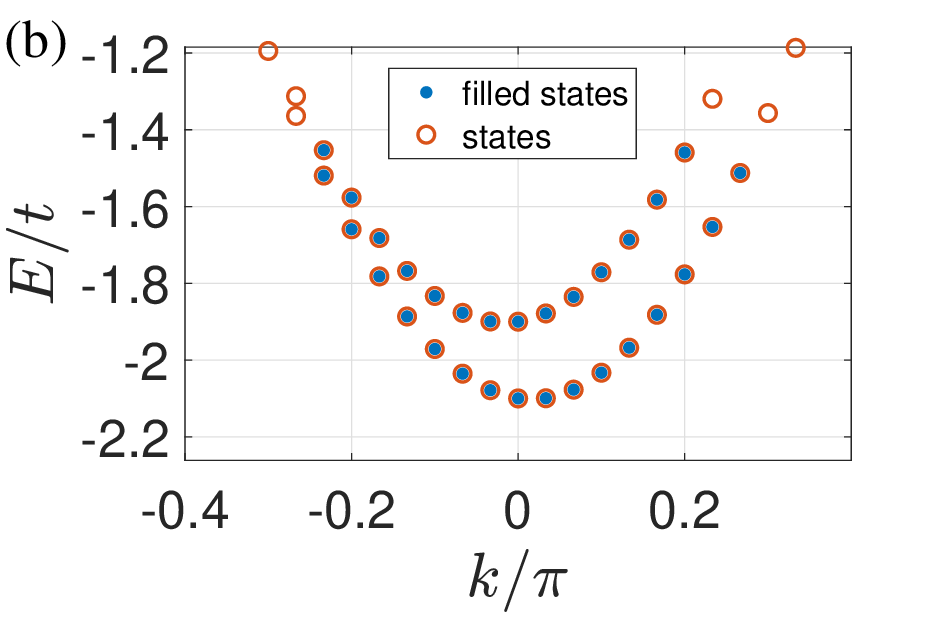}
 \caption{(a) PC versus number of electrons on a ring with $N=60$. (b) Energy versus wave number for the states (orange open circles). For quarter filling, filled states are shown with blue dot. Other parameters: $\al=0.1t$, $b=0.1t$, $\th=0$. }
 \label{fig:A}
\end{figure}

The Hamiltonian in momentum space can be written as $H_{k}=-2t\cos{k}\si_0+\al\sin{k}\si_z+b\si_{z}$, when $\th=0$. Here, we have taken the lattice spacing to be $1$. 
At non-zero values of $b$, time-reversal symmetry is broken, and for nonzero values of $\al$, parity is broken. When both $b$ and $\al$ are nonzero, time revrsal and parity are both broken which results in a net current in the ring  which is a PC. In fact, a nonzero component of the Zeeman field along $\hat z$ results in a PC in the spin-orbit coupled ring. We shall work with $\th=0$ in this paper, which means the Zeeman field is in the $\hat z$-direction.

\begin{figure*}[htb]
 \includegraphics[width=5.5cm]{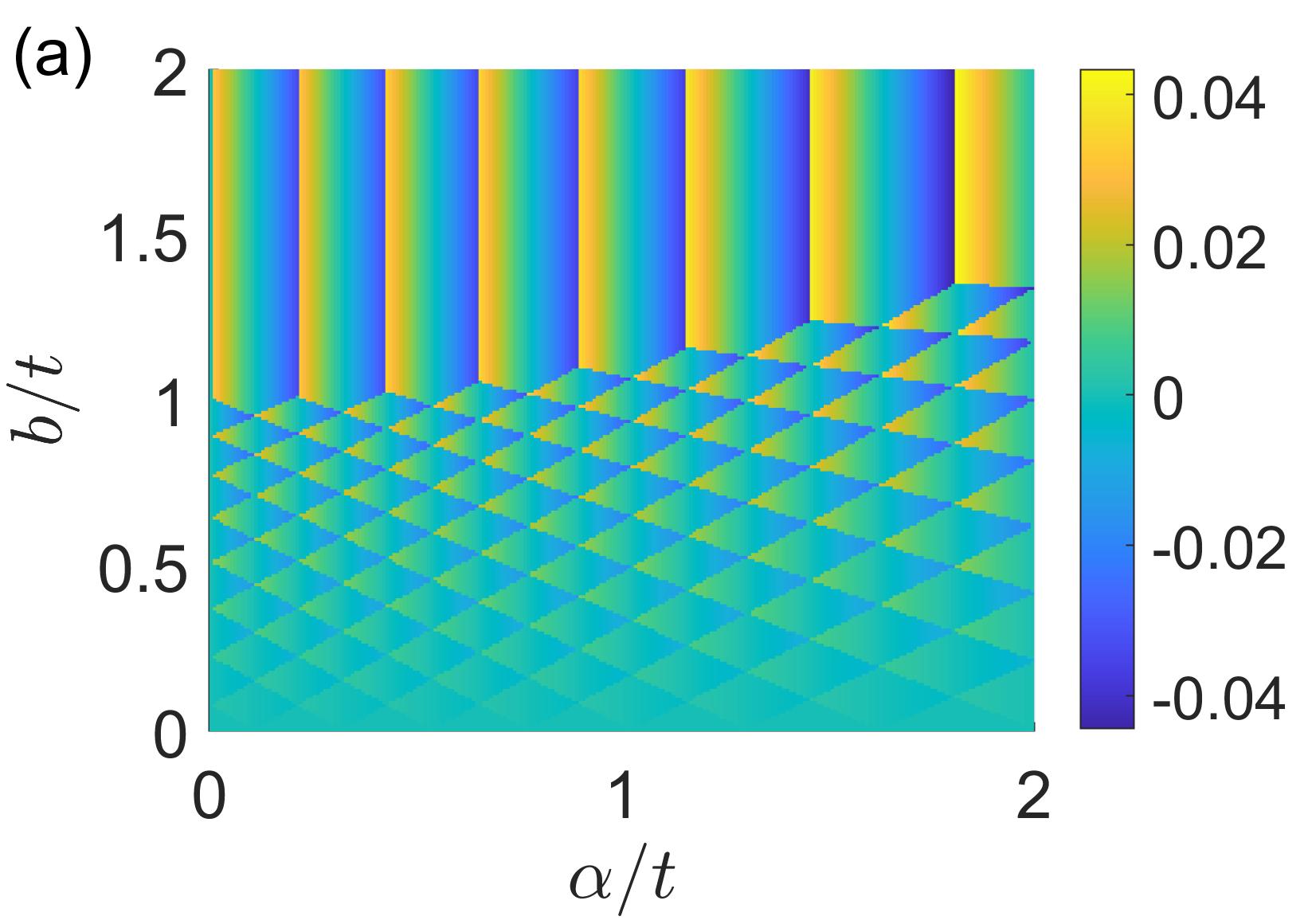}
 \includegraphics[width=5.5cm]{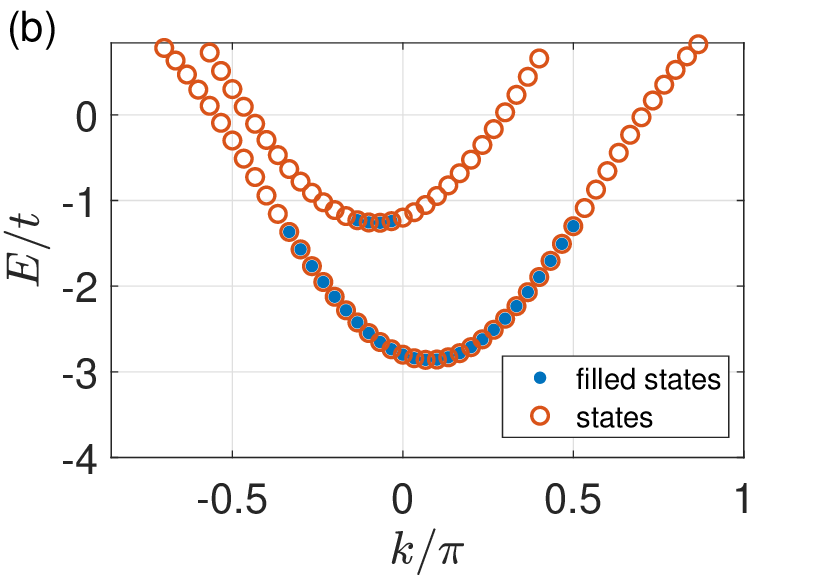}
 \includegraphics[width=5.5cm]{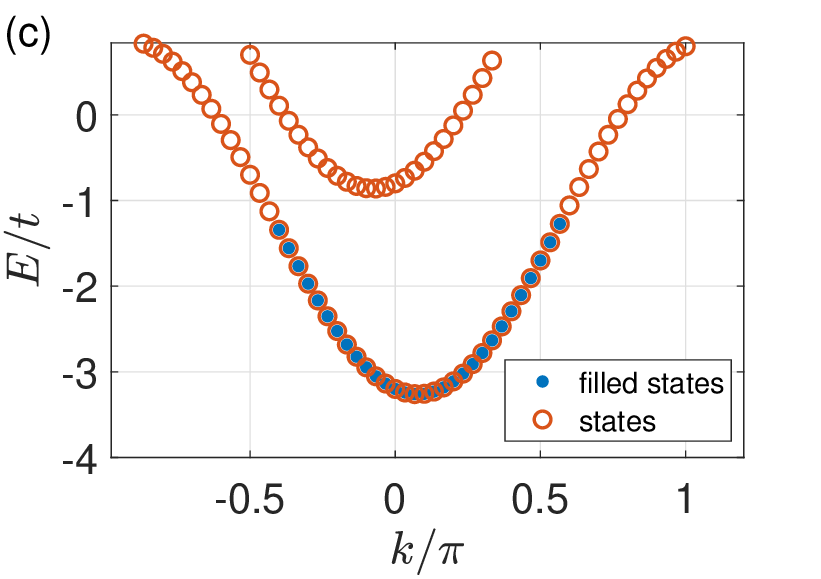}
 \caption{(a)Density plot of the PC (in units of $et/\hbar$) versus Zeeman field and the spin-orbit coupling strength.  Eigenstates and the filled eigenstates for: (b) $\al=0.5t,~ b=0.8t$ (c) $\al=0.5t, ~b=1.2t$}
 \label{fig:B}
\end{figure*}

The current carried by an individual electron between the sites $n-1$ and $n$ is given by\\
       \bea
       \label{eq:indicurrent}
       J^{(m)}_{n,n-1} &=& \f{e}{i\hbar} \Big[-t(\psi^{(m)\dg}_n \psi^{(m)}_{n-1}-{\rm c.c.}) \nn \\ && ~~+\f{i\al}{2}(\psi^{(m) \dg}_n \si_z\psi^{(m)}_{n-1}+{\rm c.c.})    \Big]\eea
       where $\psi^{(m)}_n=\begin{bmatrix}
                      \psi^{(m)}_{n,\ua} & \psi^{(m)}_{n,\da}\\       \end{bmatrix}$$^T$ is an eigenfunction at site $n$ and $e$ is the electron charge.
The index $m$ runs over different eigenstates arranged in ascending order of  energy. If $N_e$ electrons are filled, the  total current is given by
\beq 
       \label{eq:totalcurrent}
       J = \sum_{m=1}^{N_e} J^{(m)}_{n,n-1}
\eeq

Choosing $\alpha=0.1t$, $b=0.1t$, $\th=0$ and $N=60$, we diagonalize the Hamiltonian and plot the PC in the ring versus the number of electrons in Fig.~\ref{fig:A}(a) and the eigenenergies versus wave number $k$ in Fig.~\ref{fig:A}(b). 

 We find that for even number of electrons, the total current is close to zero. This can be explained using Fig.~\ref{fig:A}(b).  
  With an even number of electrons, for every right-moving electron, there is a left-moving electron with almost equal magnitude of the velocity, thus (almost) cancelling the currents pairwise. This is not true when the number of electrons is odd. The last electron filled has no partner, and it contributes significantly to the PC. Hence, contribution to the PC by the last electron is dominant for odd number of electrons, and all the filled states contribute to the PC for even number of electrons. 
 
For $\alpha=b=0$ in a ring with even number of sites, each single particle state is 4-fold degenerate since it can be labelled by $(\pm k, \pm \sigma)$. Further, the velocities of the $+k$ and $-k$ states are exactly equal and opposite. Thus, in this case, there is no current when the number of electrons is a multiple of 4 (i.e. $N_e \mod 4 = 0$) since the ground state is non-degenerate and $+k$ and $-k$ for each spin is filled. When $N_e \mod 4 = 1$ or $3$, the ground state has a degeneracy of 4 (assuming that the simultaneous eigenstates of the Hamiltonian and momentum operator are filled). Two of those states will have a non-zero current in one direction while the other two will have a current in the other direction. So, in these two cases there will be a PC. When $N_e \mod 4 = 2$, the ground state is 6 fold degenerate. Two of these states will carry a current, but in opposite directions and of twice the magnitude of the current in $N_e \mod 4 = 1$ or $3$ cases.  The remaining 4 will carry no current. So, even in the simplest case of $\alpha=b=0$, there is an effect of the number of electrons, except that it is not an odd-even effect but a $\mod 4$ effect and the current can be identically zero. Now, for $\alpha \neq 0$ but $b=0$, each single particle state is 2-fold degenerate since $(k,\sigma)$ and $(-k, -\sigma)$ are degenerate and they also have exactly equal and opposite velocities. This time there will be an odd even effect with the even filling case giving an identically zero current. For the case $\alpha \neq 0$ and $b \neq 0$, the single particle states are non-degenerate and so, in principle, there is a non-zero current for every filling. The states: $(k,\sigma)$ and $(-k, -\sigma)$  do not have exactly equal and opposite velocities. In fact, the difference is just $2\alpha\cos{k}/\hbar$. If $b$ is sufficiently small, like in the case considered here, for even filling the current is very close to zero while for odd filling it is not.

For even $N$ we find that the PCs with $N_e$ electrons filled and $2N-N_e$ electrons filled are equal in magnitude and opposite in sign. This feature can be explained by the following argument. The dispersion for the Hamiltonian in eq.~\ref{eq:Ham} is $E=-2t\cos{k}\pm(\alpha\sin{k}+b)$. Let us denote dispersions for the two bands by $E_{\pm}$. Under $k\to\pi-k$, $E_-(k)\to -E_{+}(k)$. This means that if $m$-th electron from band bottom in band $\si$ ($\si=\pm$) has wavenumber $k$, the $m$-th electron from the top of the band (enumerated in descending order) has wavenumber $\pi-k$. The velocity of electrons in the two bands are $v_{\pm}=2t\sin k\pm\alpha\cos{k}/\hbar$. This implies that $v_-(k)=v_+(\pi-k)$ which is the same as saying the currents carried by the $m$-th electron from the bottom of the bands is the same as the current carried by the $m$-th electron from the top of the band. The fully filled bands do not carry any current. So, at filling $2N-N_e$, the topmost $N_e$ electrons are missing which carry the same current as the bottom most $N_e$ electrons. This makes the currents at fillings  $N_e$ and $2N-N_e$ equal in magnitude and opposite in sign. Hence, for even $N$, the current at half-filling is zero.  Further, the allowed values of $k$ are $2\pi n/N$, where $n=0,1,2,..,N-1$. For even $N$, $k$ is allowed in addition to $\pi-k$. But this argument breaks down for odd $N$  since $\pi-k$ is not allowed when $k$ is allowed. The current at half filling need not be zero for odd $N$.

We plot the PC as a function of $\alpha$ and $b$ in Fig.~\ref{fig:B}(a) for $\th=0$ and quarter filling. We can see from Fig.~\ref{fig:B}(a) that for a fixed value of $\al$ current shows one behavior, in which it oscillates between negative and positive values as we change $b$ from zero to a critical value $b_c$. But above $b_c$, the current becomes constant as we change $b$. These two behaviors can be explained as follows. For low values of $b$, states from both the spin up and spin down bands are occupied, and this can be seen from Fig.~\ref{fig:B}(b). So, electrons from both bands contribute to the current. But for large $b$, the two bands in dispersion relation are completely separated as shown in Fig~\ref{fig:B}(c). So, if we further increase $b$, the occupied states will not change, and the bands are separated more than before. This does not change the current.

\begin{figure}[htb]
 \includegraphics[width=4.2cm]{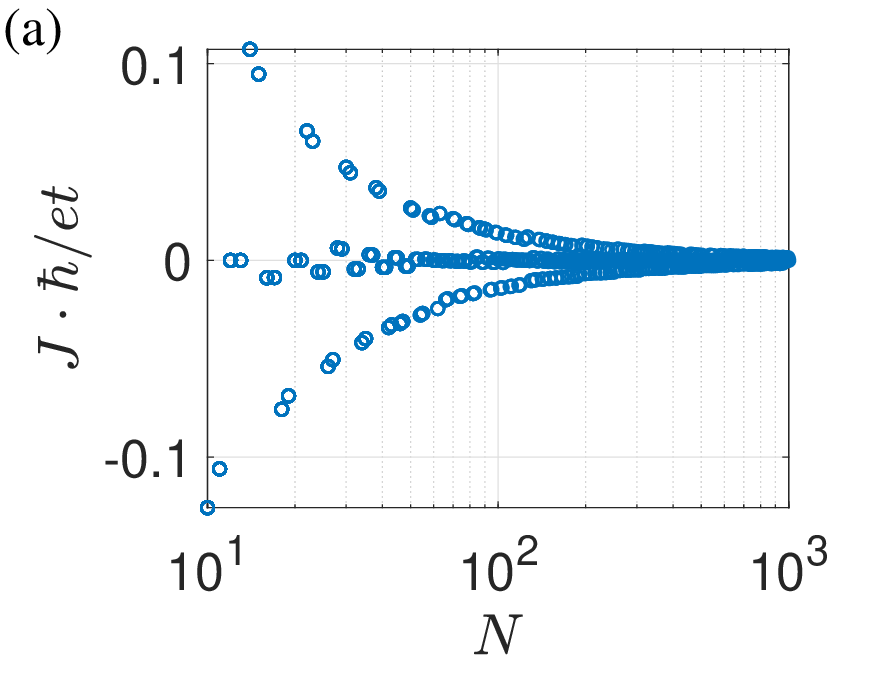}
 \includegraphics[width=4cm]{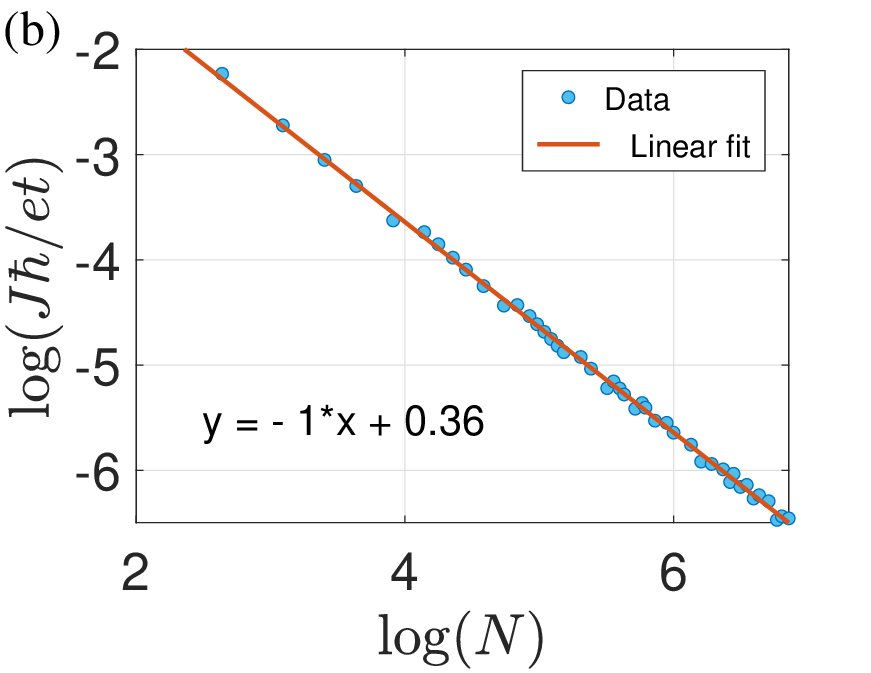}
 \caption{(a) PC versus the size of the ring at fixed filling ($N_e=[N/2]$). $\al=b=0.1t$. (b) The top envelope is plotted in the log-log scale and fitted with a linear fit. It can be seen that $J$ oscillates as $N$ increases with an envelope that decays as $1/N$.}
 \label{fig:C}
\end{figure}

In Fig.~\ref{fig:C}(a) we have plotted PC as a function of the number of lattice sites for the quarter filling case with parameters $\al=b=0.1t$ and $\th=\phi=0$. We can see that for a large value of $N$,  PC oscillates around zero with a decaying envelope. The envelope is fitted with a linear fit in the log-log plot (see Fig.~\ref{fig:C}(b)). The fit has a slope of $-1$. So, the envelope decays as $1/N$. This can be understood from the fact that PC has a dominant contribution from the single electron filled at last. The current is quadratic in $\psi$ and  $\psi$ at a site is proportional to $1/\sqrt{N}$ making the current go as $1/N$.

\begin{figure}[htb]
 \includegraphics[width=4.32cm]{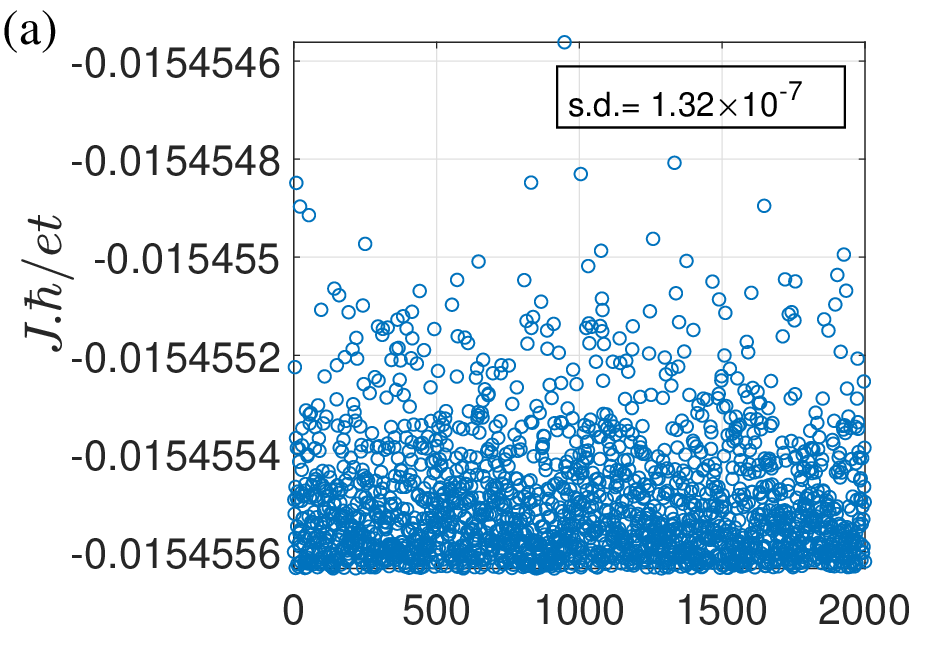}
 \includegraphics[width=4cm]{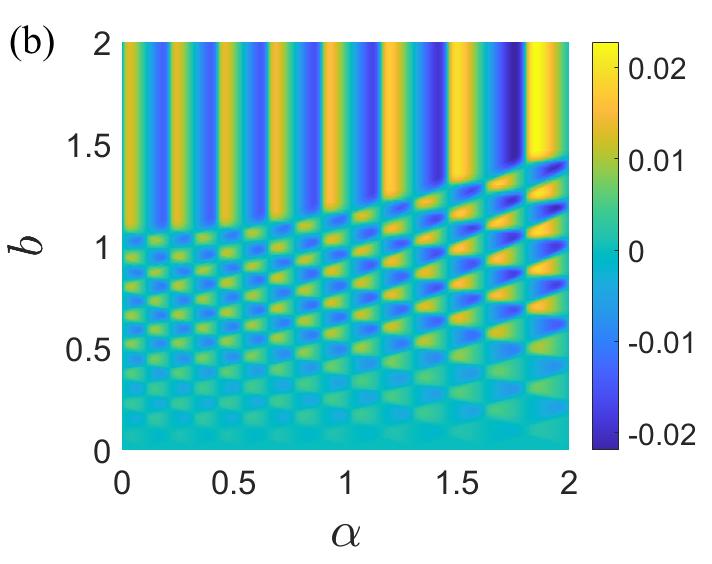}
 \caption{Effect of disorder. (a) PC for different disorder configurations for $\al=2.35\times 10^{-3}t$, $b=1.15\times 10^{-4}t$,  $w=b$, $\th=0$, $N=90$, $N_e=45$. Standard deviation (s.d.) is shown in the inset text. (b) Configuration averaged PC versus $(b,\al)$ for $w=0.5t$, $\th=0$, $N=60, ~N_e=30$. The disorder configurations are the same for each value of $(b,\al)$. Compare (b) with Fig.~2~(a). }
 \label{fig:D}
\end{figure}

\begin{figure*}[htb]
\centering
 \includegraphics[width=5.5cm]{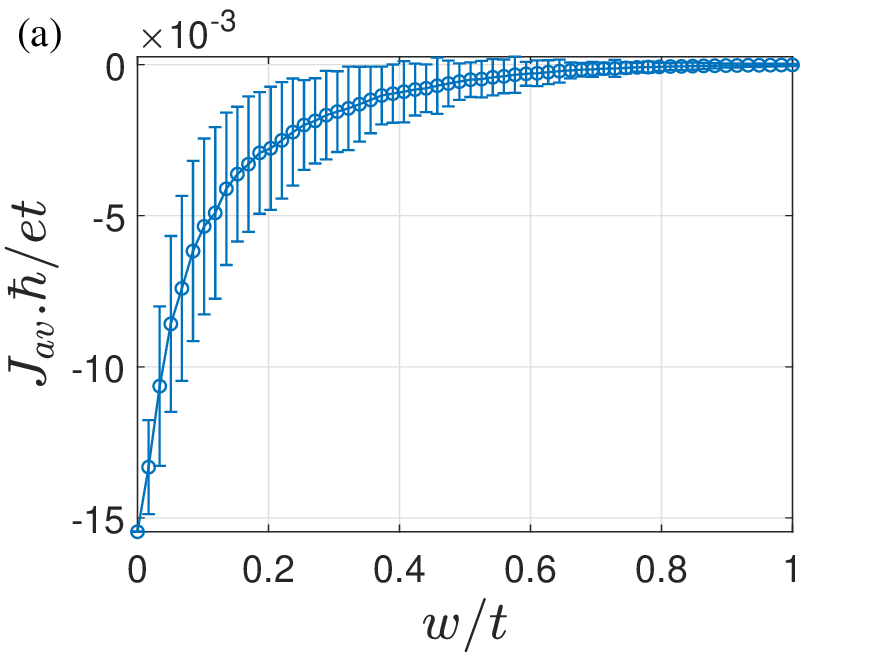}
 \includegraphics[width=5.5cm]{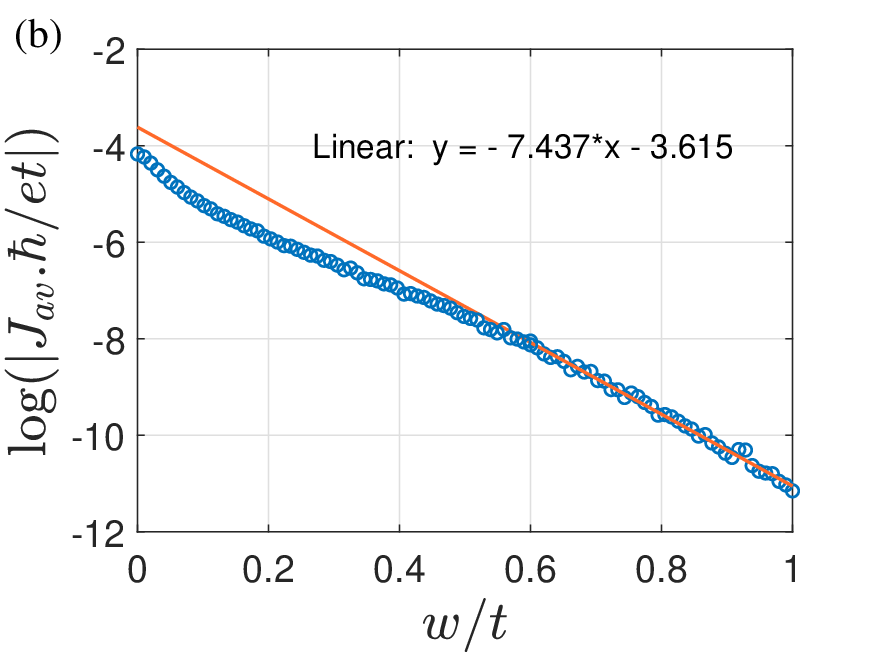}
 \includegraphics[width=5.5cm]{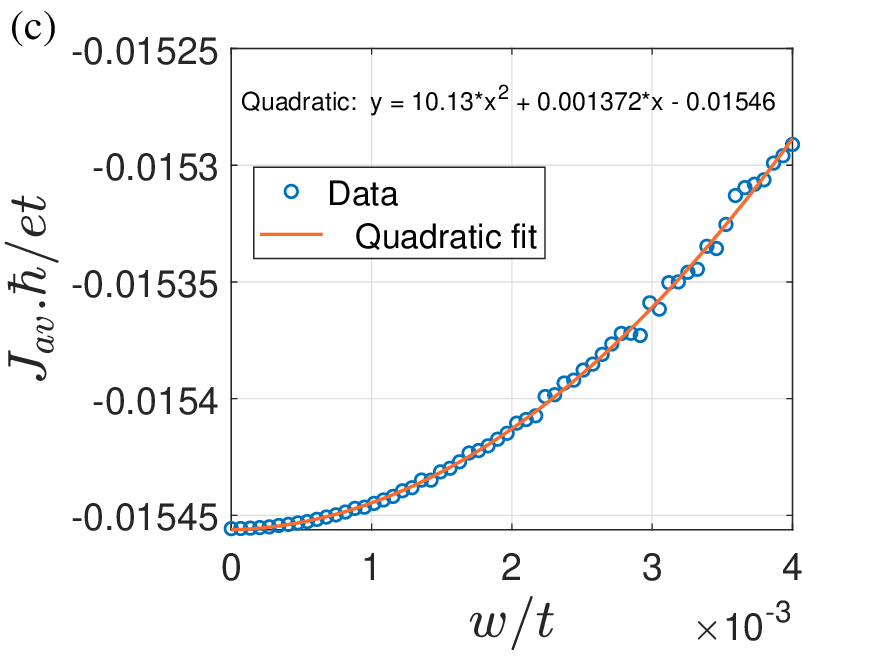}
 \caption{(a) Configuration averaged PC versus disorder strength $w$, (b) PC is fitted with a quadratic fit for weak $w$, (c) PC is plotted in log scale and fitted with a linear fit for strong $w$. All figures are at fixed filling ($N_e=[N/2]$) for $N=90$, $\al=2.35\times 10^{-3}t$, $b=1.15\times 10^{-4}t$, $\th=0$.}
 \label{fig:E}
\end{figure*}

\section{Disordered ring}
The Hamiltonian for  spin-orbit coupled ring under the influence of a Zeeman field with disorder in the on-site potential is given by
\bea
H &=& \sum_{n=1}^{N} \Big [-t(c_{n+1}^\dg c_n+h.c)+\f{\al}{2}(ic_{n+1}^\dg \si_z c_n+{\rm h.c.})\nn \\ &&+bc_{n}^\dg \si_{\th,\phi}c_n \Big]+\sum_{n=1}^{N} \ep_n c_{n}^\dg c_{n}~,  \label{eq:H-dis}
\eea
where $\epsilon_n$ is the disorder at site $n$ with $\epsilon_n$ chosen randomly in the range $[-w,w]$ (where $w$ is the strength of the disorder). We work with $\th=0$ as in the previous section. The expressions for the current carried by a single electron and the total current are the same as eq.~\ref{eq:indicurrent} and eq.~\ref{eq:totalcurrent}.

\begin{figure}[htb]
\centering
 \includegraphics[width=4.2cm]{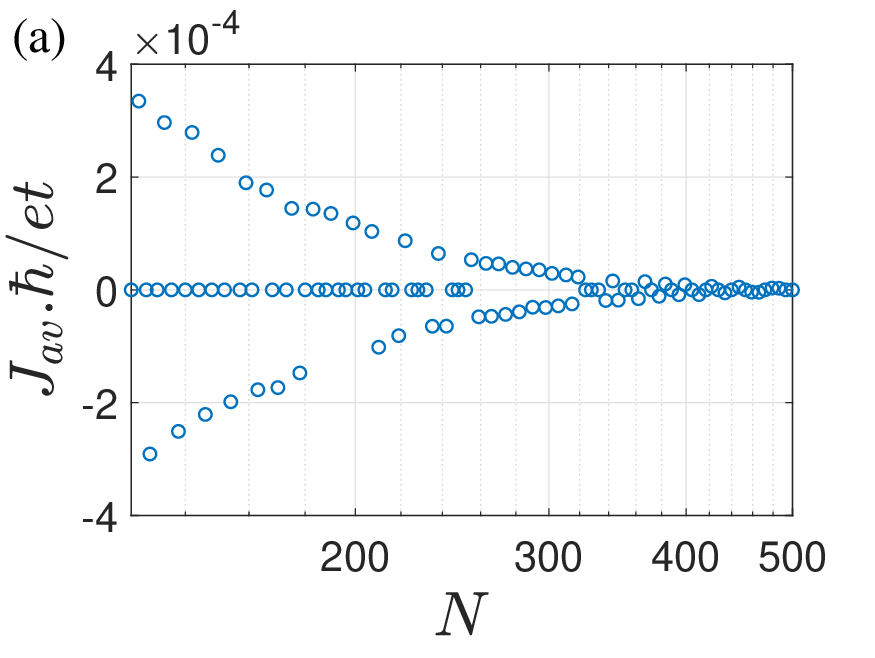}
 \includegraphics[width=4.2cm]{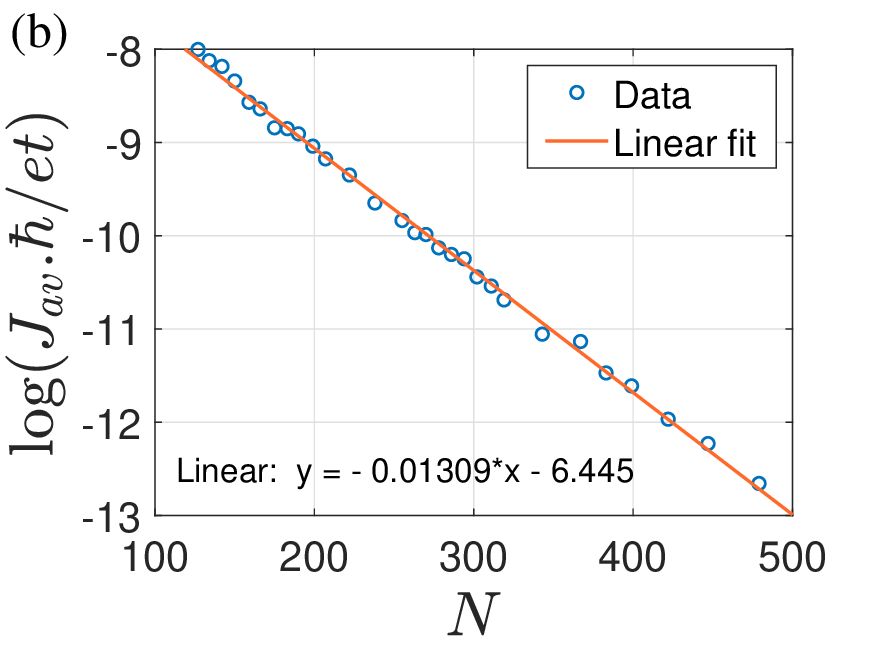}
 \caption{(a) Configuration averaged PC versus the size of the ring at fixed filling ($N_e=[N/2]$) for $w=0.5t$, $\th=0$, $\al=2.35\times 10^{-3}t$, $b=1.15\times 10^{-4}t$. (b) The top envelope is fitted in a log scale with a linear fit.}
 \label{fig:F}
\end{figure}

We plot the numerically evaluated PC for different disorder configurations in Fig.~\ref{fig:D}(a) for $\al=2.35\times 10^{-3}t$, $b=1.15\times 10^{-4}t$,  $w=b$, $N=90$, $N_e=45$. We find that the standard deviation (SD) is $1.32 \times 10^{-7}$. The above ($\al$, $b$) values are experimentally relevant for InAs nanowires~\cite{InAs2010}. Here SD is much less compared to the mean -0.0155. When the number of electrons is even, both the mean and SD of the PC are practically zero for the above choice of parameters. This is because, the energy states for $\pm k$ are almost degenerate (in absence of disorder) due to the small values of $b, \alpha$ and the currents carried by these states cancel in pairs. Adding disorder to such a Hamiltonian does not alter PC much. 
For other choices of parameters both the mean and SD of the PC may not be zero for even number of electrons. For example, if $\al=b=0.1t$, $N_c=2000$ ($N_c$ is the number of different disorder configurations) and $w=b$ are chosen for even $N_e$, the mean and the standard deviation of the PC are $1.16 \times 10^{-4}$ and $3.04 \times 10^{-5}$ respectively. In Fig.~\ref{fig:D}(b) we plotted PC versus $(b,\al)$ for $w=0.5t$, $N=60, ~N_e=30$, $N_c=1000$. Here for each value of $(b,\al)$ PC is averaged over the same set of one thousand disordered configurations. By comparing Fig.~\ref{fig:D}(b) with Fig.~\ref{fig:B}(a), we can see the value of PC decreases due to  disorder and the transition from negative to positive value of PC is not as sharp as in the ballistic case. 

We have plotted the configuration averaged PC versus strength of disorder with SD shown as error bars in Fig.~\ref{fig:E}(a) for $N=90$, $N_e=[N/2]$, $\al=2.35\times 10^{-3}t$, $b=1.15\times 10^{-4}t$, $N_c=1000$. We can see that the PC goes to zero along with the standard deviation as $w$ increases. We have plotted the configuration averaged PC on a log-scale in Fig.~\ref{fig:E}(b) along with a linear fit for the range $w=0.6t$ to $w=t$. For strong $w$ the PC varies exponentially with $w$. But for weak $w$, the scenario is different. In Fig.~\ref{fig:E}(c) we can see the PC is fitted with a quadratic fit for weak $w$. So, PC varies as $w^2$ for weak $w$.

In Fig.~\ref{fig:F}(a) we plotted configuration averaged PC versus the size of the ring for $w=0.5t$, $\al=2.35\times 10^{-3}t$, $b=1.15\times 10^{-4}t$, $N_c=1000$ and $N_e=[N/2]$. We can see from Fig.~\ref{fig:F}(a) that the magnitude of oscillation goes to zero as $N$ increases. Here the value of the PC is low as compared to the ballistic case (see Fig.~\ref{fig:C}(a)). In Fig.~\ref{fig:C}(b) we saw that the envelope of the PC varies as $1/N$, but here it is not like that. We have plotted the envelope of Fig.~\ref{fig:F}(a) in log-scale in Fig.~\ref{fig:F}(b) and fitted with a linear fit. So, for strong $w$, the PC varies exponentially with $N$. But for weak $w$, the PC varies exponentially only for large $N$.

\begin{figure}[htb]
    \includegraphics[width=4.2cm]{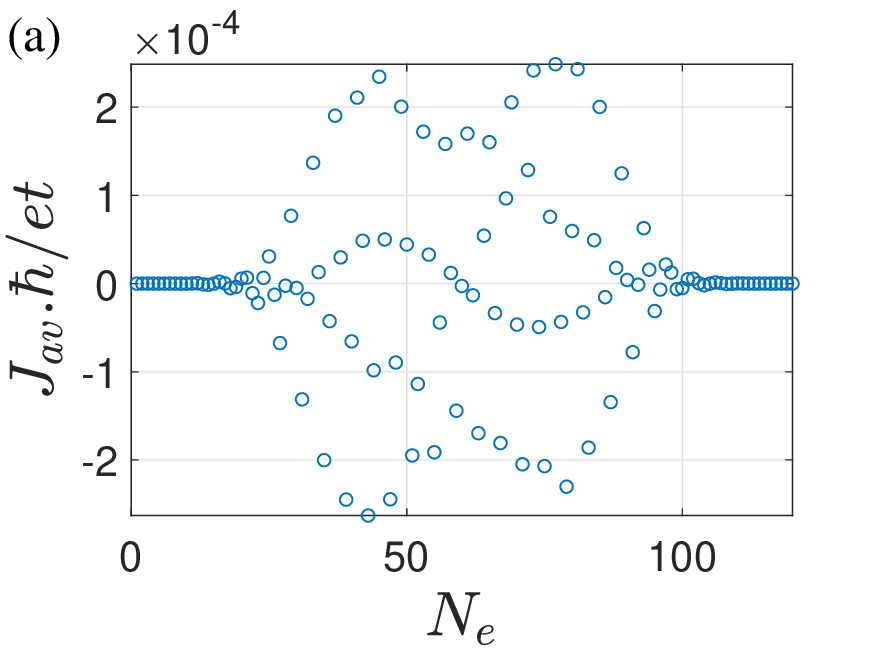}
    \includegraphics[width=4.2cm]{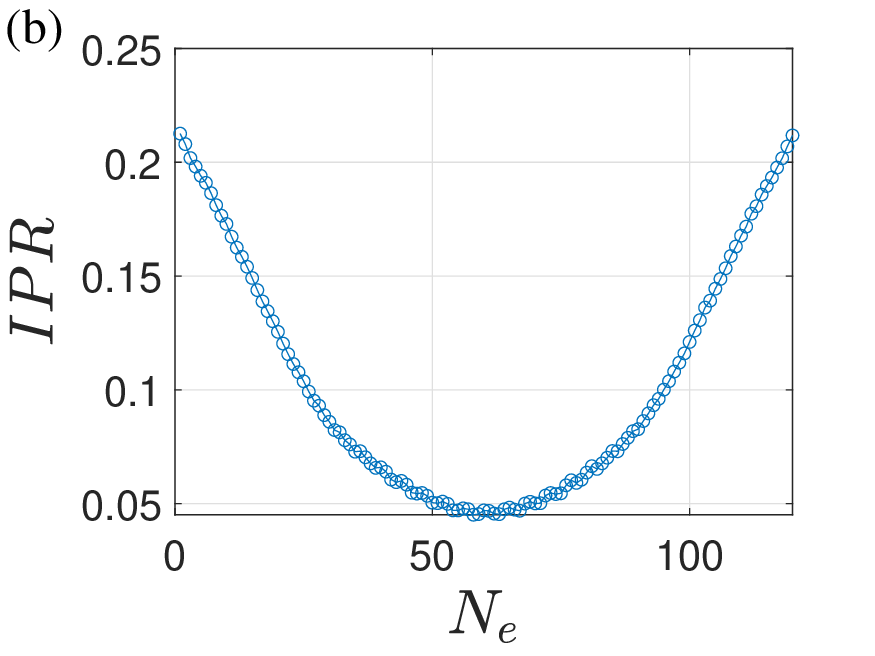}
    \caption{(a) Configuration averaged PC versus number of electrons.(b) Inverse participation ratio (IPR) of the last electron versus the number of electrons. In all figures: $N=60$, $w=t$, $\th=0$, $\al=b=0.1t$,$N_c=10000$. }
    \label{fig:G}
\end{figure}

 We have seen the behavior of the PC with electron filling in the ballistic case (see in Fig.~\ref{fig:A}(a)). Now it is interesting to know the behavior of  PC versus filling for the disordered case. In Fig.~\ref{fig:G}(a) we have plotted the configuration averaged  PC of the ring versus the number of electrons for $N=60$, $w=t$, $\al=b=0.1t$ and $N_c=10000$. Here  PC is averaged over ten thousand disordered configurations. We can see from Fig.~\ref{fig:G}(a) that for low filling the PC is zero, but for moderate filling we have a significant PC value. This can be explained using Fig.~\ref{fig:G}(b). In Fig.~\ref{fig:G}(b) we have plotted the inverse participation ratio (IPR) of the last electron versus the number of electrons for the same set of parameters. Here IPR is averaged over $N_c=10000$ disordered configurations. We can see from Fig.~\ref{fig:G}(b), for low filling IPR is high compared to moderate filling. That means the last electron is strongly localized in the low-filling case compared to moderate filling. So the current is suppressed due to the localization which leads to zero current in the low filling case.
 
\begin{figure}[htb]
    \includegraphics[width=6cm]{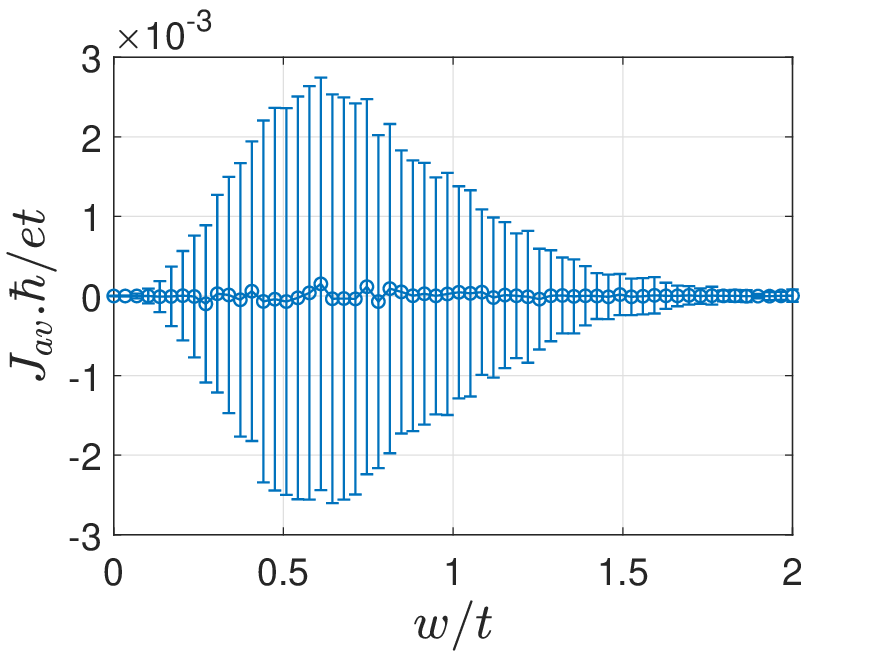}
    \caption{Configuration averaged PC versus the strength of disorder at fixed filling ($N_e=[N]$) for $N=90$, $\al=b=0.1t$, $\th=0$, $N_c=1000$.}
    \label{fig:H}
\end{figure}
In the ballistic case, we saw that for even $N$ at half-filling, the PC is zero. However, with the disorder, the configuration averaged PC is zero at half filling for even $N$, while the standard deviation increases with $w$, attains a maximum, and goes to zero for very strong $w$. We can see this from Fig.~\ref{fig:H} where configuration averaged PC is plotted versus $w$ for $\al=b=0.1t$, $\th=0$, $N=90$, $N_e=[N]$ and $N_c=1000$. Here we can see for $w=0$ both the configuration averaged PC and standard deviation are zero. As $w$ increases the configuration averaged PC remains zero, while the standard deviation increases with $w$, reaches a maximum, and then starts decreasing with a further increase in $w$. For very strong $w$, the standard deviation also goes to zero. In experiments for a given sample, the disordered configuration is fixed, and depending on the strength of the disorder the PC of the sample may not be zero at half-filling. But it lies anywhere between the negative to the positive value of SD as shown in the error bar in Fig.~\ref{fig:H}. Though PC is zero in the ballistic case at half-filling, the disorder helps in getting a nonzero PC. 

\section{Noncollinear Zeeman field}
\begin{figure}
\includegraphics[width=4cm]{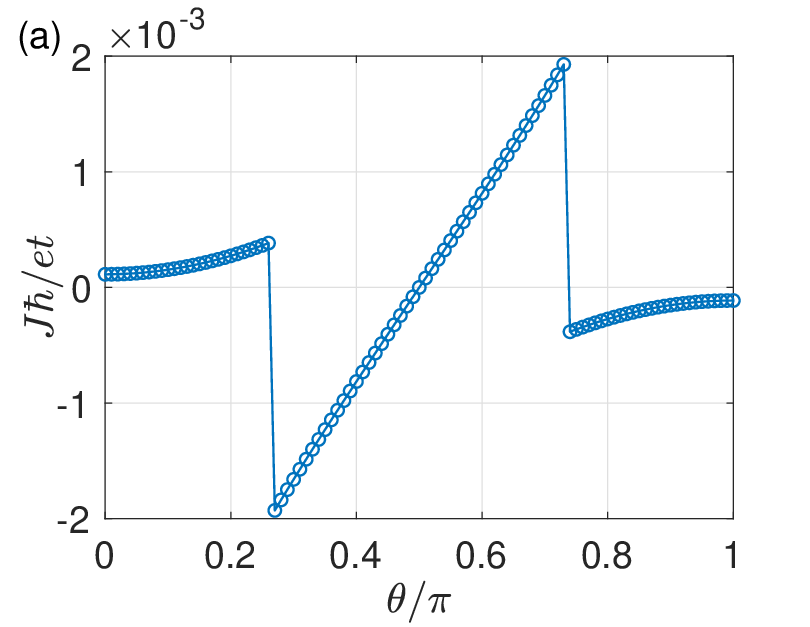}
\includegraphics[width=4cm]{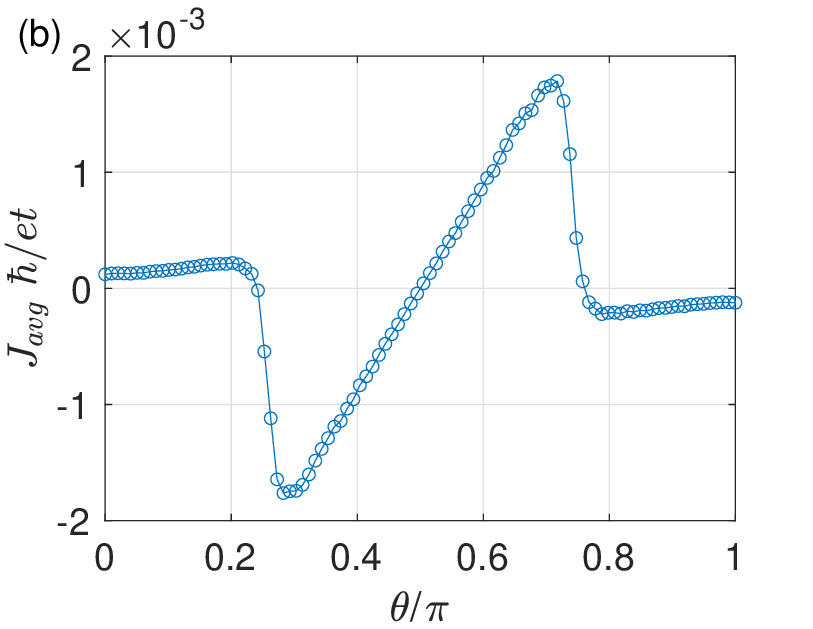}
\includegraphics[width=4cm]{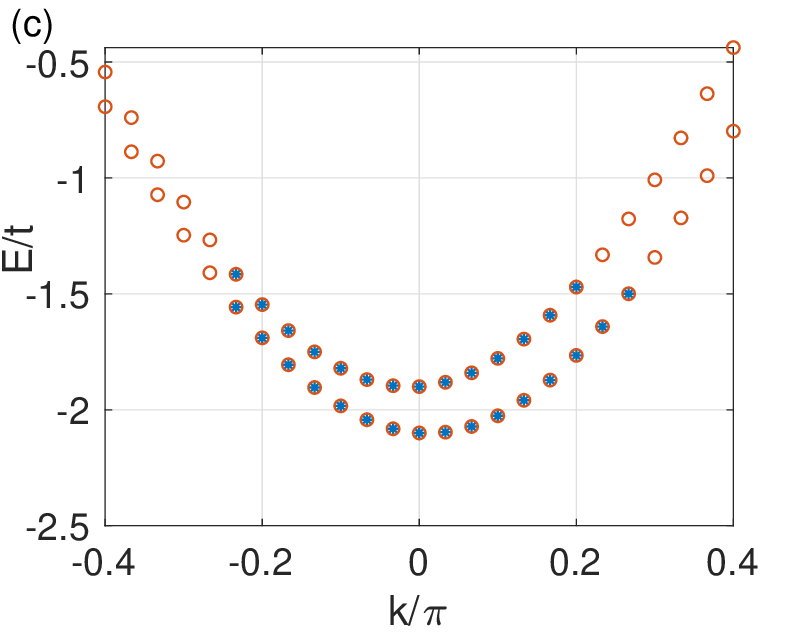}
\includegraphics[width=4cm]{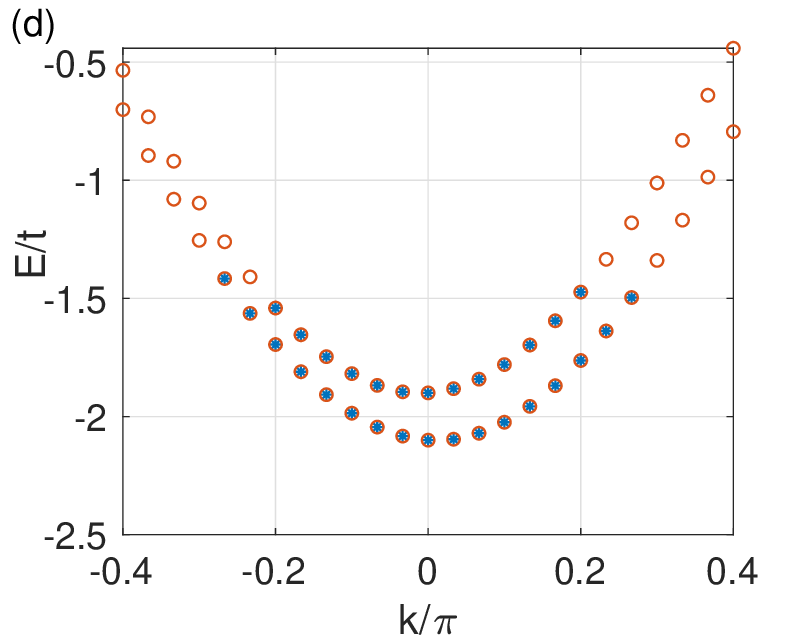}
\caption{PC versus $\th$ for clean (a) and disordered (b) rings. Disorder strength $w=0.2t$ and $N_c=1000$ for (b). Eigenenergy versus wavenumber for the clean case for $\th=0.25\pi$ (c) and $\th=0.28\pi$ (d). Open circles in represent unoccupied eigenstates while the circles with dot at the centre represent occupied states. Parameters: $\al=b=0.1t$, $N=60$, $N_e=[N/2]$.  }\label{fig:PC-th}
\end{figure}

In this section we study behavior of PC when the Zeeman field is not in the same direction as the spin-orbit field. In eq.~\eqref{eq:Ham}, $(\th,\phi)$ define the direction of Zeeman field. PC does not depend on $\phi$. This can be shown by transforming the Hamiltonian by unitary operator $U=e^{-i\phi\sigma_z/2}$. In Fig.~\ref{fig:PC-th}(a), we plot the dependence of PC on $\th$ for $\al=b=0.1t$, $N=60$, $N_e=30$ for a clean ring. One feature of this plot is that PC obeys $J(\th)=-J(\pi-\th)$. This is because, under $\th\to\pi-\th$ accompanied by $\phi\to\phi+\pi$, $b\to-b$ and the Hamiltonian becomes time reversal partner of the original Hamiltonian under $b\to-b$. Under time-reversal, the current reverses its direction.  Another feature of the plot is that around $\th=0.27\pi$, there is a sudden jump  in the value of PC as $\th$ is varied. This is because, the last electron suddenly shifts the band it occupies as $\th$ is changed and the velocities in the two bands are quite different. This can be seen from Fig.~\ref{fig:PC-th}(c,d) where the energy eigenstates are plotted for $\th=0.25\pi$~(c) and $\th=0.28\pi$~(d). In Fig.~\ref{fig:PC-th}(b), PC is plotted for same set of parameters, but with disorder having a strength $w=0.2t$. It can be seen that the plot for the case of clean ring gets smoothened with a slight variation in values of PC.   

\section{Persistent spin currents}

\begin{figure}[htb]
    \includegraphics[width=6cm]{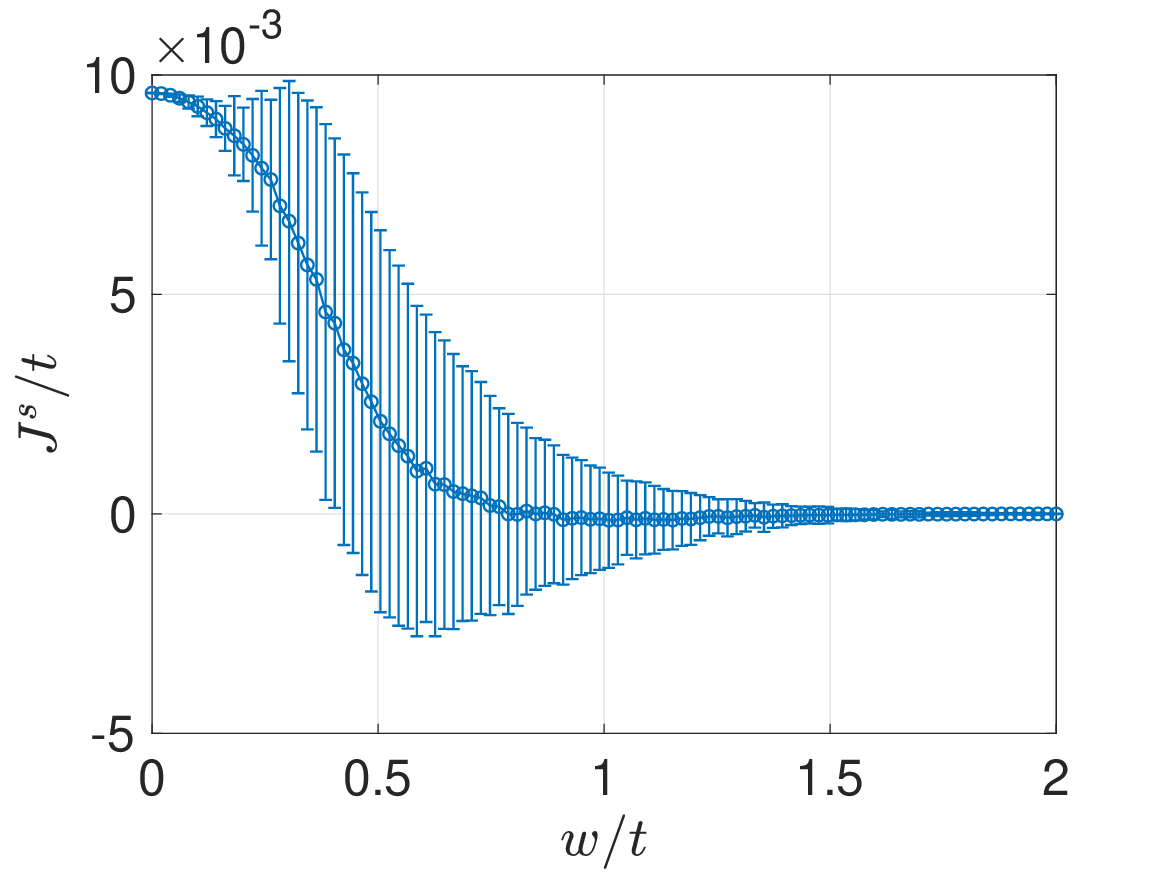}
    \caption{Configuration averaged PSC (error bar shows standard deviation) versus the strength of disorder at fixed filling ($N_e=[N]$) for $N=90$, $\al=b=0.1t$, $\th=0$, $N_c=1000$.}
    \label{fig:I}
\end{figure}
The Hamiltonian [Eq.\eqref{eq:H-dis}] commutes with $\si_z$ when $\th=0$. Hence, the spin $\si_z$ is a conserved quantity. The spin current corresponding to the spin density defined by $\rho_s=(\hbar/2)\psi^{\dagger}\si_z\psi$ can be obtained by using continuity equation. The expression for spin current  is given by 
\bea
J^{s,(m)}_{n,n-1} &=& \f{1}{2i} \Big[-t(\psi^{(m)\dg}_n \si_z\psi^{(m)}_{n-1}-{\rm c.c.}) \nn \\ && ~~+\f{i\al}{2}(\psi^{(m) \dg}_n \psi^{(m)}_{n-1}+{\rm c.c.})  \Big], \eea
where $m$ denotes the eigenstate of the Hamiltonian and $(n,n-1)$ is the bond on which the spin current is evaluated. The total spin current is $J^s=\sum_{m=1}^{N_e}J^{s,(m)}_{n,n-1}$. We find that for the ballistic case ($w=0$), the  persistent spin current (PSC) is inversely proportional to the system size $N$ at a fixed filling fraction. In the disordered ring, PSC varies quadratically with disorder strength for small $w$ before decaying exponentially at large $w$. 
With system size $N$, the disorder averaged PSC oscillates very much like PC with an envelope that decays exponentially. 
Disorder averaged PSC as a function of filling exhibits features similar to Fig.~\ref{fig:G}(a). A feature that distinguishes PSC from PC is that at half-filling (for even $N$), PSC is not zero in both the ballistic limit  and with disorder. In Fig.~\ref{fig:I}, we plot disorder averaged PSC versus disorder strength $w$. For large $w$, average PSC and the standard deviation in PSC both approach zero.   In fact, spin-orbit interaction is sufficient to generate a nonzero PSC without the need for a Zeeman field. 

\section{Discussion}
Spin-orbit interaction leads to an equilibrium spin current  in a quantum ring~\cite{sun2007}. This spin current can be converted to persistent charge current  by introducing magnetization to the ring~\cite{Zhang2011}. This conversion is a result of broken time-reversal symmetry, which makes the unequal flow of  currents carried by the spin-up and spin-down components of the spin current, results in a net persistent charge current in the system. Our work is same in spirit as that of ref.~\cite{Zhang2011}. An important difference between ref.~\cite{Zhang2011} and our work is that, while the direction of spin orbit field changes with every bond in their work, it is uniform in our work. For a radial spin-orbit field, the Zeeman field also should be radial to get PC. Applying a uniform Zeeman field does not result in PC in their model whereas a uniform Zeeman field with a component along the spin-orbit field can produce PC in our case.  Applying a uniform  Zeeman field is experimentally possible by proximitising the ring to a ferromagnetic insulator whereas applying a nonuniform Zeeman field is extremely difficult experimentally. Also, ref.~\cite{Zhang2011} studies PC in ballistic rings, whereas our work includes  the effect of disorder. 

Previous works have discussed the effect of spin-orbit coupling on persistent currents in mesoscopic rings threaded by a magnetic flux~\cite{fuji1993,moskalets2000,ding2007}.
Fujimoto and Kawakami~\cite{fuji1993} study the interplay between Hubbard interaction and spin-orbit coupling in mesoscopic ring threaded by a flux exactly using Bethe-ansatz method to explore the effects on orbital magnetism of the ring. Moskalets~\cite{moskalets2000} studied the effects of spin-orbit coupling and Zeeman field on interacting ring threaded by a flux, described by Bosonization. Ding and Dong~\cite{ding2007} explored the effect of spin orbit coupling on PC in a ring threaded by a magnetic flux connected to an Anderson impurity.  
Our work is different from these in the sense that we show the appearance of  persistent currents  in spin-orbit coupled rings under an applied Zeeman field with no magnetic flux threading the ring. Electron transport in spin-orbit coupled rings is accompanied by an Aharonov-Casher phase factor~\cite{avishai2019}. The rings described in our work is a special case of the rings studied by Avishai et. al~\cite{avishai2019}, where the electric field points radially inwards, and the effective magnetic field (spin-dependent) is uniform across the ring. Moreover, we have an additional Zeeman field in our study. While the work by Avishai et. al. focuses on conductance across a ring connected to two leads, we study the persistent currents in isolated rings in this work. 

Next, we discuss parameter values that are relevant to experimental setups. For InAs, the parameters are given as follows~\cite{InAs2010,kleemans2007,ahn2021}: \( t = 7.225 \, \text{eV} \), \( \alpha = 17 \, \text{meV} \), \( a = 0.4 \, \text{nm} \) (lattice spacing), \( b = 0.836 \, \text{meV} \),  \( w \sim 1 \, \text{meV} \) and a size of about \( 36 \text{nm}\). These correspond to the ratios \( \alpha = 2.35 \times 10^{-3}t \), \( b = 1.15 \times 10^{-4}t \), \( w = b \), and a size \( N = 90 \). This parameter set is used in our study, as shown in Fig.~\ref{fig:D}, Fig.~\ref{fig:E}, and Fig.~\ref{fig:F}. Under these conditions, we obtain a current of \( -0.0154 \, et/\hbar = -27 \, \mu \text{A} \), which is significantly large. This is same in order of magnitude as the current in a ring having same values of parameters  threaded by a magnetic flux in absence of Zeeman and spin-orbit fields (experimentally measured PC in rings~\cite{P} having a circumference of $2{\rm \mu m}$ threaded by a flux has a magnitude of $1{\rm nA} $). However, increase in  ring size or  disorder strength diminishes the current, reducing its practical magnitude.

\section{Summary and Conclusion}
Our work focusses on PC in spin-orbit coupled rings under the influence of Zeeman field in contrast with PC in rings threaded by a magnetic flux. PC in our setup requires nonzero values of spin-orbit coupling and Zeeman field. Our investigation reveals that PC in a ballistic ring is inversely proportional to  the system size. For even number of sites, PC is zero at half filling.  The dependence of PC on $(b,\al)$ exhibits two distinct behaviors for large and small $b$. Below a critical value of $b$, PC oscillates with $b$ for a fixed $\al$ and above the critical value of $b$ PC remains constant.   On-site disorder leads to suppression of PC typically. PC decays with disorder exponentially for large strength of disorder and quadratically for small disorder. PC decays exponentially with system size for large system sizes. The critical system size above which PC decays exponentially with system size is smaller for strong disorder and large for weak disorder.  Interestingly, for even number of sites, PC is zero at half filling for the ballistic ring, but disorder can enhance PC in a single sample, though the configuration averaged PC is zero. The standard deviation of PC increases with disorder strength, reaches a maximum, and then decreases to zero at large strength of disorder. These findings not only contribute to a deeper understanding of mesoscopic quantum transport phenomena, but also have practical implications for the development of quantum devices and technologies. 

\acknowledgements 
We thank Adhip Agarwala, Udit Khanna and Dhavala Suri for useful discussions.  AS and BKS  thank Science and Engineering Research Board (India) Core Research grant (CRG/2022/004311) for financial support.  AS thanks University of Hyderabad  for funding through Institute of Eminence PDF. BKS thanks Ministry of Social Justice and Empowerment, Government of India for fellowship through NFOBC. 

\bibliography{ref_phe}

\end{document}